\documentclass[twocolumn,showpacs,amsmath,amssymb]{revtex4}

\usepackage{graphicx}
\usepackage{dcolumn}
\usepackage{bm}

\newcommand{\so}{SO(N_c)}
\newcommand{\su}{SU(N_c)}
\newcommand{\spto}{Sp(2N_c)}
\newcommand{\dgc}{d({\cal G})}
\newcommand{\lgc}{\ell({\cal G})}
\newcommand{\cuv}{c_{UV}}
\newcommand{\cir}{c_{IR}}
\newcommand{\beq}{\begin{eqnarray}}
\newcommand{\beqa}{\begin{eqnarray*}}
\newcommand{\eeq}{\end{eqnarray}}
\newcommand{\eeqa}{\end{eqnarray*}}

\begin{document}

\title{Patterns of chiral symmetry breaking and a candidate for a $C$-theorem
  in four dimensions}

\author{Jesper Levinsen}
\email{levi@nbi.dk}
\affiliation{The Niels Bohr Institute\\ Blegdamsvej 17, DK-2100 Copenhagen
{\O}\\ Denmark}

\date{October 26, 2001}

\begin{abstract}
  We test a candidate for a four-dimensional $C$-function. This is done by
  considering all asymptotically free, vectorlike gauge theories with $N_f$
  flavors and fermions in arbitrary representations of any simple Lie group.
  Assuming spontaneous breaking of chiral symmetry in the infrared limit and
  that the value of the $C$-function in this limit is determined by the
  number of Goldstone bosons, we find that only in the case of a theory with
  two colors and fermions in one single pseudo-real representation of $SU(2)$
  the $C$-theorem seems to be violated. Conversely, this might also be a sign
  of new constraints, restricting the number of flavors consistent with
  spontaneous chiral symmetry breaking. For all other groups and
  representations we find that this candidate $C$-function decreases along
  the renormalization group flow.
\end{abstract}

\maketitle

\section{Introduction}

For two-dimensional field theories, Zamolodchikov's $C$-theorem \cite{zamo}
states that it is possible to construct a $C$-function which decreases
monotonically along the renormalization group (RG) flow. This is viewed as a
sign of irreversibility of the RG flow.  This $C$-function depends on the
couplings of the theory and is stationary at conformal fixed points. At these
it reduces to the central charge of the theory, and since this may be
interpreted as a measure of the degrees of freedom, the $C$-theorem
demonstrates the loss of information from short-distance behaviour to the
infrared limit.

Since a $C$-theorem measuring irreversibility of RG flows holds in two
dimensions, it is natural to ask whether this is also true in four
dimensions. A proof that this is indeed the case has been proposed
\cite{forte} but it has not been entirely accepted \cite{osborn,cap01}. In
four dimensions we face the problem that Zamolodchikov's two-dimensional
$C$-function does not generalise in a unique way. In fact there are three
possible generalisations as well as combinations of those. In ref.
\cite{phd}, Cardy's proposal for a $C$-function in four dimensions
\cite{cardy} has been discussed (see also
[7-12]). It was seen, that the
inequality $\cuv\geq\cir$ was satisfied for all combinations compatible with
asymptotic freedom under the same assumptions as we will use below. Cardy's
proposed $C$-function is constructed from the Euler term in the trace of the
energy-momentum tensor.  However, in this paper we will test an alternative
candidate for a $C$-function, given by the coefficient proportional to the
Weyl tensor in the trace anomaly \cite{duff}.  The properties of this
$C$-function has been investigated previously. In a study of ${\cal N}=1$
supersymmetric gauge theories it was concluded \cite{anselmi2} that no linear
combination of this $C$-function together with Cardy's proposal is decreasing
in all models (except the trivial combination consisting of only Cardy's
function). Also, perturbative studies around perturbative fixed points have
shown that in some cases this $C$-function increases along the flow
\cite{cap91}. However, the only non-supersymmetric gauge theory which have
been studied non-perturbatively is QCD \cite{shore} where it was found that
this $C$-function actually does decrease along the flow. Although there is
evidence pointing towards Cardy's proposal, it is still of interest to study
the properties of all possible candidates. We will follow the procedure of
\cite{phd}.

The theories we are considering are asymptotically free, vector-like gauge
theories with $N_f$ massless Dirac-fermions.  Our assumption is that the
$C$-function is given by the same expression also when we approach the
infrared limit, where chiral symmetry is assumed spontaneously broken. We
will not consider exotic scenarios where, in this limit, there are other
massless states than precisely those required by Goldstone's theorem. We also
assume that there are no fundamental scalars in the theory.  Then, for a
theory with $N_f$ flavors of fermions the one-loop $\beta$-function takes the
form
\beq
\beta (g) = -\frac{g^3}{16\pi^2}\left[ \frac{11}{3}\ell({\cal G}) -
  \frac{4}{3}\ell(r)N_f \right]+\dots,
\eeq
and since we require that the theory be asymptotically free
\beq
N_f < \frac{11}{4}\frac{\lgc}{\ell(r)}.
\label{asymptotic}
\eeq
Here $\lgc$ and $\ell(r)$ are the indices of the representations of the gauge
bosons (${\cal G}$) and fermions ($r$), respectively. This bound may be too
weak since asymptotic freedom and chiral symmetry breaking may be lost for
smaller values of $N_f$.

The number of flavors $N_f$ and the representation $r$ is seen to be
constrained by the demand that the theory be asymptotically free.  We now
want to compare the values of the $C$-function in the ultraviolet and infrared
fixed points, for all irreducible representations of compact, simple Lie
groups and for all $N_f$ compatible with asymptotic freedom.

The $C$-function we want to test takes the value \cite{duff}
\beq
c = N_0+6N_{1/2}+12N_1
\eeq
at the fixed points. Here, $N_0$ is the number of massless real scalars,
$N_{1/2}$ is the number of massless Dirac fermions and $N_1$ is the number
of massless gauge bosons. In the ultraviolet limit we have
\beq
c_{UV} = 6N_f d(r) + 12\dgc
\eeq
where $d(r)$ and $\dgc$ are the dimensions of the representation $r$ and
of the gauge group $\cal G$, respectively. 

The value of the $C$-function at the infrared fixed point is given by the
number of massless degrees of freedom, i.e. the dimension of the Goldstone
manifold. Thus it depends on the way in which chiral symmetry is
spontaneously broken. There are believed to be three ways in which this can
happen \cite{peskin}, a conjecture which has very recently been
investigated in the context of lattice gauge theories
\cite{patterns}. The three classes of breaking to consider are:

\begin{itemize}
\item{The representation of the fermions is pseudo-real.
    Spontaneous breaking of chiral symmetry in this case is expected to break
    $SU(2N_f)$ to $Sp(2N_f)$.}
\item{The representation is complex. Here we expect the symmetry breaking
    pattern to be $SU(N_f)\times SU(N_f)\rightarrow SU(N_f)$.}
\item{The representation is real. This case is similar to the pseudo-real
    case, but here the expected symmetry breaking pattern is
    $SU(2N_f)\rightarrow SO(2N_f)$ }
\end{itemize}

These cases can be labelled by their Dyson-indices, $\beta=1$, $\beta=2$ and
$\beta=4$ respectively, due to a connection to Random Matrix Theory
\cite{jv}. That these classes of spontaneous symmetry breaking actually do
occur in the $N_c\rightarrow\infty$ limit has been proven
in ref. \cite{cw} for the
classes $\beta=2$, $\beta=4$ with arguments easily extended to $\beta=1$
\cite{phd}. Each case assumes maximal symmetry breaking consistent with the
Vafa-Witten theorem \cite{vw} and thus gives an upper bound on the value of
the $C$-function in the infrared. Determining the number of Goldstone bosons,
we have
\beq
c_{IR} = \left\{
  \begin{array}{lcc}
    N_f(2N_f-1)-1 & \mbox{for} & \beta=1,\\
    N_f^2-1 & \mbox{for} & \beta=2, \\
    N_f(2N_f+1)-1 & \mbox{for} & \beta=4.
  \end{array}
\right.
\eeq

It is useful to note that $\cir(\beta=2)\leq\cir(\beta=1)<\cir(\beta=4)$.
This leads to the following lemma \cite{phd}:

\vspace {5mm}
{\em Lemma}: If $\cuv\geq\cir(\beta[r_0])$ for fermions in a representation
$r_0$ of ${\cal G}$ with dimension $d(r_0)$ and index $l(r_0)$, then this
will also hold for all other representations $r$ of ${\cal G}$ with
$\cir(\beta[r])\leq\cir(\beta[r_0])$, $d(r)\geq d(r_0)$ and $l(r)\geq l(r_0)$.

\vspace{5mm} It is in the sense of the last two inequalities, that we will
talk about the smallest representation.

\section{Groups and representations}

We now perform a systematic investigation of simple, compact Lie-groups
in the Cartan classification.

\vspace{5mm}
\begin{figure}
  \includegraphics[width=.9\columnwidth]{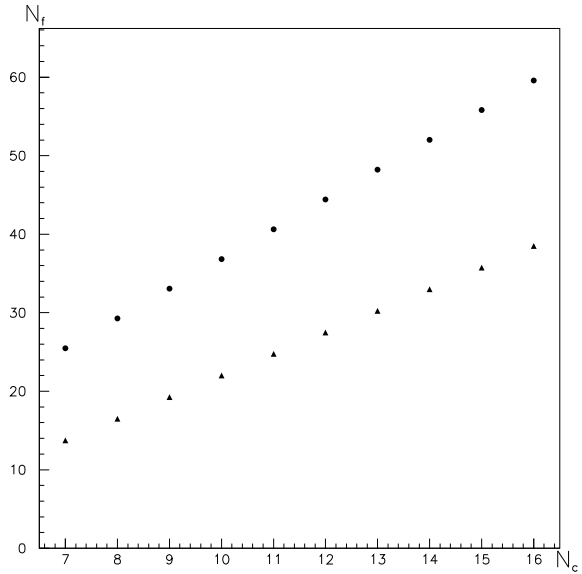}
  \caption{\label{son}Upper bounds on the number of flavors as a function of
    the number of colors for the defining representation of
    $SO(N_c)$, requiring $\cuv\geq\cir$ ($\bullet$) and asymptotic freedom
    ($\blacktriangle$).}
\end{figure}
$\so$: Since $SO(2)$ is abelian, $SO(3)\sim SU(2)$, $SO(5)\sim Sp(4)$,
$SO(6)\sim SU(4)$ and since $SO(4)$ is not simple, here we only consider
$\so$, $N_c\geq7$. The dimension of $\so$ is $\dgc=N_c(N_c-1)/2$ while
the index of the adjoint representation is $\lgc=N_c-2$.

The defining representations are real ($\beta=4$) and for these $\ell(r)=1$
and $d(r)=N_c$. Thus
\beq
\cuv & = & 6 N_c N_f+6 N_c(N_c-1) \nonumber  \\
\cir & = & N_f (2 N_f+1)-1.
\eeq
Solving the condition $\cuv\geq\cir$ with respect to $N_f$ we find, that
\beq
N_f \leq \frac{1}{4} \left( 6 N_c - 1 + \sqrt{3(28 N_c^2 - 20N_c + 3)}\right),
\eeq
which is well above the bound from the condition of asymptotic freedom
\beq
N_f < \frac{11}{4}(N_c-2).
\eeq
This is illustrated in Fig. \ref{son}.
We now use the lemma on all other representations of $\so$, since all have
$\ell(r)\geq1$ and $d(r)\geq N_c$. Thus, this function satisfies the
$C$-theorem for the group $\so$

\vspace{5mm}
$\spto$: The dimension of $\spto$ is $\dgc=N_c(2N_c+1)$ and the index of
the adjoint representation is $\lgc=2(N_c+1)$.

\begin{figure}
  \includegraphics[width=.9\columnwidth]{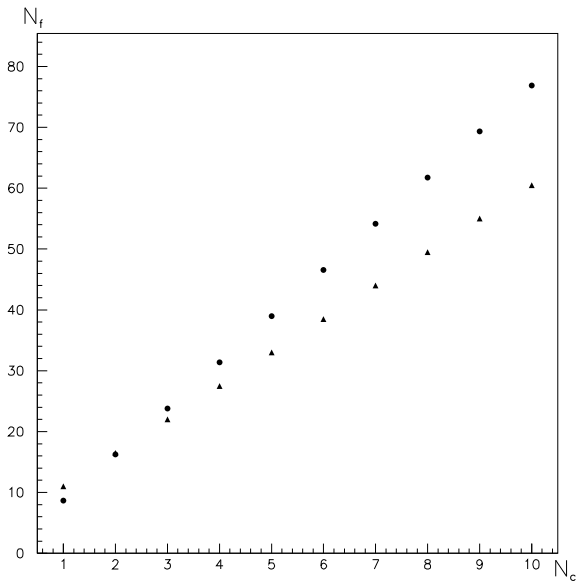}
  \caption{\label{spps}The upper bounds on $N_f$ for
    the fundamental representation of $Sp(2N_c)$,
    requiring $\cuv\geq\cir$ ($\bullet$) and asymptotic freedom
    ($\blacktriangle$).}
\end{figure}

The representations of $\spto$ are all real or pseudo-real.
The fundamental representations of $\spto$ are all pseudo-real and have
$\ell(r)=1$ and $d(r)=2N_c$. Thus
\beq
\cuv & = & 12 N_c (2 N_c+1) + 12 N_c N_f \nonumber\\
\cir & = & N_f (2 N_f-1)-1.
\eeq
Again we require that $\cuv\geq\cir$ which corresponds to
\beq
N_f \leq \frac{1}{4}\left(12N_c+1+\sqrt{3(112N_c^2+40N_c+3)}\right).
\eeq
This should be compared to the condition of asymptotic freedom
\beq
N_f < \frac{11}{2}(N_c+1).
\eeq
This is illustrated in Fig. \ref{spps}.
It is seen that for $N_c=1$ corresponding to $Sp(2)$, we have a non-trivial
bound, $N_f\leq8$ instead of the bound of $N_f<11$ obtained from demanding
asymptotic freedom. For $N_c=2$ it looks like there is another non-trivial
condition but in this case both conditions give the same bound, namely
$N_f\leq16$.
Since all pseudo-real representations have $\ell(r)\geq1$ and $d(r)\geq2N_c$
we conclude from the lemma that the $C$-theorem is fulfilled for all
pseudo-real representations as long as $N_c\geq2$.

\begin{figure}
  \includegraphics[width=.9\columnwidth]{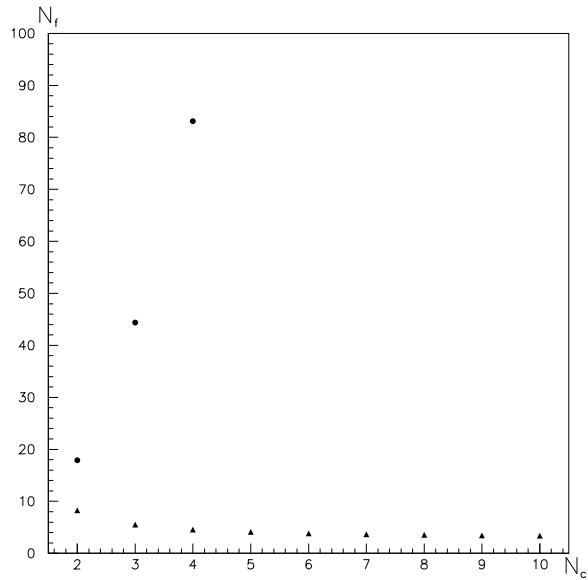}
  \caption{\label{spr}Upper bounds on $N_f$ for
    the smallest real representation of $Sp(2N_c)$,
    requiring that the $C$-theorem be fulfilled ($\bullet$)
    and asymptotic freedom ($\blacktriangle$).}
\end{figure}

Considering now the real representations, the smallest of these has
$d(r)=N_c(2N_c-1)-1$ and $\ell(r)=2(N_c-1)$ (note that this representation
is trivial for $Sp(2)$). In this case
\beq
\cuv & = & 12 N_c (2 N_c+1) \nonumber\\
&&+ 6 N_f(N_c ( 2 N_c-1)-1) \nonumber\\
\cir & = & N_f (2 N_f+1)-1.
\eeq
The condition $\cuv\geq\cir$ is now equivalent to
\beq
\!\! N_f & \leq &
\frac{1}{4} \left( 12N_c^2 - 6N_c -7 + \sqrt{3}\nonumber \right.\\
&& \left.\times\sqrt{48N_c^4-48N_c^3+20N_c^2+60N_c+19} \right).
\eeq
The condition (\ref{asymptotic}) ensuring asymptotic freedom becomes
\beq
N_f < \frac{11}{4}\frac{N_c+1}{N_c-1}
\eeq
and the situation is illustrated in Fig. \ref{spr}. It is seen, that
for $N_c\geq 2$ we have $\cuv\geq\cir$ and since this was the smallest real
representation, by the lemma we conclude that the $C$-theorem is valid
for all real representations of $Sp(2N_c)$.

\begin{figure}
  \includegraphics[width=.9\columnwidth]{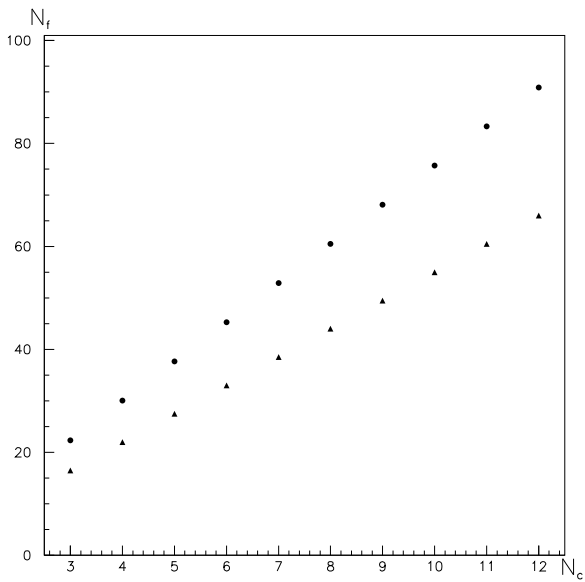}
  \caption{\label{suc}The upper limits on $N_f$ as a function of $N_c$ for
    the fundamental representation of $\su$,
    requiring $\cuv\geq\cir$ ($\bullet$) and asymptotic freedom
    ($\blacktriangle$).}
\end{figure}

\vspace{5mm}
$\su$: The group $SU(2)$ plays a special role, so we begin by considering
$N_c\geq3$. For $\su$ we have $\dgc=N_c^2-1$ and $\lgc=2N_c$. The fundamental
representations are complex, they all have $\ell(r)=1$ while the dimension is
$d(r)=N_c$. The $C$-function in the two limits thus becomes
\beq
\cuv & = & 12(N_c^2-1) + 6 N_c N_f \nonumber\\
\cir & = & N_f^2-1
\eeq
which leads to
\beq
N_f \leq 3 N_c + \sqrt{-11 + 21 N_c^2}.\label{sufundamental}
\eeq
The condition (\ref{asymptotic}) becomes $N_f<11N_c/2$, so that
(\ref{sufundamental}) is automatically satisfied as illustrated in Fig.
\ref{suc}. This result was already noticed in ref. \cite{shore}.
All other complex representations have larger dimensions and indices
than the fundamental representation. Thus we conclude, that the $C$-theorem
is valid for all complex representations of $\su$.

The adjoint representations are always real and have $\ell(r)=2N_c$ and
$d(r)=N_c^2-1$. In this case
\beq
\cuv & = & 12( N_c^2-1) + 6( N_c^2-1) N_f \nonumber\\
\cir & = & N_f (2 N_f+1)-1.
\eeq
Thus
\beq
N_f \leq \frac{1}{4}\left(6 N_c^2 -7+ \sqrt{3(12 N_c^4+4N_c^2-13)}\right).
\eeq
This is much above the bound of
$N_f<11/4$ from the requirement of asymptotic freedom, and thus
$\cuv\geq\cir$ for the adjoint representations of $\su$.

But there are other real and pseudo-real representations, and for these it is
not possible to select the smallest. Thus it seems like we have to check the
representations on a case-by-case basis. This was done in ref. \cite{phd} up
to and including $SU(9)$ and it was found, that only $SU(4)$ and $SU(6)$ have
representations that are both smaller than those of the adjoint
representations and satisfy the constraint from asymptotic freedom.  For
$SU(4)$ there are three relevant real representations, out of which the
smallest has $\ell(r)=2$ and $d(r)=6$. For this representation we get
\beq
\cuv & = & 36 N_f+180 \nonumber\\
\cir & = & N_f (2 N_f+1)-1,
\eeq
and thus
\beq
N_f \leq \frac{1}{4}(35+9\sqrt{3}) 
\eeq
which is above the bound of $N_f<11$ from the condition
(\ref{asymptotic}). We now use the lemma on the other
two real representations.

$SU(6)$ only has one relevant representation which is pseudo-real and has
$\ell(r)=6$ and $d(r)=20$. The condition $\cuv\geq\cir$ is automatically
satisfied, since it translates into $N_f\leq (121 + 3\sqrt{2001})/4$ which is
much above the bound of $N_f < 11/2$ from (\ref{asymptotic}).

Turning now to the group $SU(2)$ the smallest pseudo-real representation is
the fundamental, which has $\ell(r)=1$ and $d(r)=2$. The bound of $N_f<11$
from the requirement of asymptotic freedom is seen to be above the bound of
$N_f\leq (13 + \sqrt{465})/4\approx8.6$. However, this is as expected since
$SU(2)\sim Sp(2)$ so this is in fact the same representation which had this
feature for $Sp(2)$. For the real representations the smallest has
$\ell(r)=4$, $d(r)=3$ so that the bound of $N_f\leq (17 + 3 \sqrt{65})/4$
from the requirement of $\cuv\geq\cir$ is much above the bound of $N_f<11/4$
from (\ref{asymptotic}).

\vspace{5mm}
{\em The exceptional groups:} For the exceptional groups we only
need to calculate the bound on $N_f\ell(r)$ from the requirement of
asymptotic freedom and then check the possible representations. This is done
based on the tables of \cite{wgm}.

$E_6$ has adjoint index 24 and the condition (\ref{asymptotic}) thus becomes
$N_f\ell(r)<66$. The relevant representations are the fundamental and the
adjoint. The fundamental representation is complex and has $d(r)=27$,
$\ell(r)=6$ and the adjoint has $d(r)=78$.  In both cases $\cuv\geq\cir$.

In $E_7$ the adjoint index is 36 and thus $N_f\ell(r)<99$ from
(\ref{asymptotic}). There are two relevant representations, the fundamental
is pseudo-real and has $d(r)=56$ and $\ell(r)=12$, while the adjoint
representation has dimension $d(r)=133$. Again we find that $\cuv\geq\cir$.

$E_8$ only has one relevant representation, the fundamental which coincides
with the adjoint. Here $d(r)=248$ while $\ell(r)=60$, and again it is seen
that $\cuv\geq\cir$.

For the group $F_4$ the adjoint index is $\ell({\cal G})=18$ and the
condition for asymptotic freedom is $N_f\ell(r)<99/2$. Again there are two
relevant representations. The fundamental representation is real and has
$d(r)=26$ and $\ell(r)=6$ while the adjoint has $d(r)=52$. By the lemma,
since $\cuv\geq\cir$ for the fundamental representation this is also the case
for the adjoint.

The last group, $G_2$, has adjoint index $\lgc=8$ and thus $N_f\ell(r)<22$.
There are three relevant representations, all real.  The adjoint
representation has $d(r)=14$ while the fundamental is the smallest with
$d(r)=7$, $\ell(r)=2$. Since $\cuv\geq\cir$ for the fundamental we need not
check the others.

\section{Conclusion}

To conclude, we have performed a systematic investigation of all
representations of simple, compact Lie groups. We have shown that in only one
case, the requirement of asymptotic freedom is insufficient to ensure that
the candidate $C$-function considered in this paper decreases from the
ultraviolet to the infrared. This one case is for one pseudo-real
representation of the gauge group $SU(2)\sim Sp(2)$. In all other cases,
demanding asymptotic freedom alone guarantees that the inequality
$\cuv\geq\cir$ is fulfilled.

This result supports the general belief that Cardy's proposed $C$-function is
the most likely candidate if Zamolodchikov's $C$-theorem is to be extended to
four dimensions. Alternatively, this may be viewed as an indication of a
non-trivial upper limit on the number of flavors in the $SU(2)$ theory
consistent with spontaneous breaking of chiral symmetry.

\begin{acknowledgments}
The author would like to thank P. H. Damgaard for discussions.
\end{acknowledgments}


\begin{thebibliography}{X}

\bibitem{zamo}
A.~B.~Zamolodchikov,
JETP Lett.\ {\bf 43} (1986) 730.

\bibitem{forte}
S.~Forte and J.~I.~Latorre,
Nucl.\ Phys.\ B {\bf 535} (1998) 709
[hep-th/9805015]; hep-th/9811121.

\bibitem{osborn}
H.~Osborn and G.~M.~Shore,
Nucl.\ Phys.\ B {\bf 571} (2000) 287
[hep-th/9909043].

\bibitem{cap01}
A.~Cappelli, R.~Guida, and N.~Magnoli,
hep-th/0103237.

\bibitem{phd}
R.~D.~Ball and P.~H.~Damgaard,
Phys.\ Lett.\ B {\bf 510} (2001) 341
[hep-th/0103249].

\bibitem{cardy}
J.~L.~Cardy,
Phys.\ Lett.\ B {\bf 215} (1988) 749.

\bibitem{ij}
I.~Jack and H.~Osborn,
Nucl.\ Phys.\ B {\bf 343} (1990) 647.

\bibitem{cap91}
A.~Cappelli, D.~Friedan, and J.~I.~Latorre,
Nucl.\ Phys.\ B {\bf 352} (1991) 616.

\bibitem{cap00}
A.~Cappelli, G.~D'Appollonio, R.~Guida, and N.~Magnoli,
hep-th/0009119.

\bibitem{bast}
F.~Bastianelli,
Phys.\ Lett.\ B {\bf 369} (1996) 249
[hep-th/9511065].

\bibitem{anselmi1}
D.~Anselmi, D.~Z.~Freedman, M.~T.~Grisaru, and A.~A.~Johansen,
Nucl.\ Phys.\ B {\bf 526} (1998) 543
[hep-th/9708042].

\bibitem{anselmi2}
D.~Anselmi, J.~Erlich, D.~Z.~Freedman, and A.~A.~Johansen,
Phys.\ Rev.\ D {\bf 57} (1998) 7570
[hep-th/9711035].

\bibitem{duff}
M.~J.~Duff,
Nucl.\ Phys.\ B {\bf 125} (1977) 334.

\bibitem{shore}
G.~M.~Shore,
Phys.\ Lett.\ B {\bf 256} (1991) 407.

\bibitem{peskin}
M.~E.~Peskin,
Nucl.\ Phys.\ B {\bf 175} (1980) 197.

\bibitem{patterns}
P.~H.~Damgaard, U.~M.~Heller, R.~Niclasen, and B.~Svetitsky,
hep-lat/0110028.

\bibitem{jv}
J.~Verbaarschot,
Phys.\ Rev.\ Lett.\ {\bf 72} (1994) 2531
[hep-th/9401059].

\bibitem{cw}
S.~Coleman and E.~Witten,
Phys.\ Rev.\ Lett.\ {\bf 45} (1980) 100.

\bibitem{vw}
C.~Vafa and E.~Witten,
Nucl.\ Phys.\ B {\bf 234} (1984) 173.

\bibitem{wgm}
W.G. McKay and J. Patera, {\em Tables of Dimensions, Indices, and Branching
Rules for Representations of Simple Lie Algebras} (Marcel Dekker Inc.,
New York 1981).

\end{thebibliography}
\end{document}